\documentclass[aps,twocolumn,epsfig,rotate,showpacs]{revtex4}
\usepackage{epsfig}
\usepackage{graphicx}
\usepackage{amsmath}

\newcommand{\beq}{\begin{eqnarray}}
\newcommand{\eeq}{\end{eqnarray}}
\begin{document}
\title{Controlled Quantum State Transfer in a Spin Chain
}
\author{Jiangbin Gong$^{1}$
and Paul Brumer$^{2}$}
\affiliation{
$^{1}$Department of Physics and Center for Computational Science
and Engineering, \\ National University of Singapore, 117542,
Republic of Singapore\\
$^{2}$Chemical Physics Theory Group and Center for Quantum Information
and Quantum Control, \\ University of Toronto, Toronto, Canada  M5S 3H6}

\begin{abstract}
Control of the transfer of quantum information encoded in quantum
wavepackets moving along a spin chain is demonstrated.
Specifically, based on a relationship with control in a paradigm
of quantum chaos, it is
shown that wavepackets with slow dispersion can automatically emerge from
a class of initial superposition states involving only a few spins, and
that arbitrary unspecified travelling wavepackets can be nondestructively 
stopped and later relaunched with perfection. 
The results
establish an interesting application of quantum chaos studies
in quantum information science.
\end{abstract}
\date{OCT. 25, 2006}
\maketitle

\section{introduction}
Great effort is being devoted to studies of spin chains as promising
``quantum wires"  for quantum information transfer. With spin chains, a
quantum state  can be transferred without requiring an interface between
the communication channel and a quantum computer \cite{Bose03a}, {\it
i.e.}, quantum information can be transferred and processed with the same
hardware. Spin chains also allow for quantum computing with an always-on
interaction \cite{Bose03b,Twamley06}, even in the presence of a a global
control field  \cite{Twamley06}.  The latest experimental progress on
fabrication and characterization of atomic spin chains was reported in
Ref. \cite{cyrus}.  Spin chain Hamiltonians may be also realized by
atomic gas in an optical lattice.

Perfect state transfer in spin chains might occur under special
circumstances  \cite{ekert04etc,Kayetc,feder}.
However, in the general case, dispersion effects often
degrade the transmission fidelity and improving the fidelity  
becomes a central issue. 
Notably, it has been proven that
the transmission fidelity can be significantly improved
if the receiver stores the received signal in
a large quantum memory before decoding \cite{Giovannetti06}.
Another general
approach to high-fidelity quantum state transfer
advocates the use of quantum wavepackets to encode
the quantum state of a qubit \cite{Osborne,Haselgrove}. This
approach is important
because dispersion of
wavepackets can be insignificant.
In particular, Osborne and Linden \cite{Osborne} have shown that
high transmission fidelity can be achieved by
exploiting, if attainable, a Gaussian wavepacket whose shape
is well preserved. The slow dispersion of a wavepacket can be further
suppressed by applying a static parabolic (hence global)
magnetic field \cite{sun}.

In the context of the wavepacket approach to quantum state transfer, we focus
below on two questions: (1) how can one create spin
wavepackets with certain desired features, and (2) how can one control the
motion of a quantum wavepacket in a spin chain so that the packet can be
stopped at an arbitrary time, held, and then restarted later,
without loss of quantum information. Such type of controlled quantum state
transfer, if possible, should be a highly valuable tool in a variety of
situations, {\it e.g.}, cases in which the information receivers need
additional waiting time to repair a quantum memory, or to prepare for a
time window of high transmission fidelity. 
The importance of stoppable quantum state transfer may be
also appreciated by noting the analogy to the potential impact of
the stopping of light \cite{stoplight} in quantum information science.
Further,
a working scenario for the
stopping and perfect relaunching of quantum state evolution of a spin chain
should be also of considerable interest in the context of perfect quantum
state reconstruction and perfect quantum state storage in systems of
interacting
qubits \cite{giampaolo}.

In this paper we first show that
by optimizing a particular transport property using
quantum superposition states comprising
only a few spins (e.g., four or five),
wavepacket pairs with some highly desired features
emerge automatically from the ensuing dynamics.
We then demonstrate that by applying a sequence of pulsed parabolic
magnetic fields one can manipulate these wavepackets, stopping them and
later relaunching the travelling wavepackets without individually
addressing the spins. As shown below, the stopping, followed by
relaunching, can in principle perfectly preserve the quantum information
being transferred. This is made possible by taking advantage of powerful
relationships between controlling spin dynamics and
controlling quantum diffusion
dynamics
in a paradigm of quantum chaos. 

This paper is organized as follows. In Sec. II we introduce a mapping between
a Heisenberg spin chain kicked by a parabolic magnetic field and
a paradigm in quantum chaos \cite{Bose06,boness}. In Sec. III we propose
a conceptually
simple approach to the creation
of spin wavepacket pairs
moving along the spin chain with slow dispersion and other desired features.
The key result of this work is in Sec. IV, where
stopping and relaunching spin wavepackets
are studied both numerically and analytically.
Section V concludes this paper.

\section{Heisenberg Spin Chain in a Pulsed Magnetic Field and the Detla-Kicked Rotor}
Consider then an open-ended
 Heisenberg chain of $N$ spins in a constant magnetic
 field $B$ and subject to a parabolic $\delta$-pulsed magnetic field. The
 Hamiltonian is given by
 \begin{eqnarray}
 H&=&-\frac{J}{2}\sum_{n=1}^{N-1}{\bf \sigma}_{n}\cdot {\bf \sigma}_{n+1} -B\sum_{n=1}^{N}\sigma_{n}^{z}\nonumber \\
 &&+
 \sum_{j}\delta(t-jT_{0})\sum_{n=1}^{N}\sigma_{n}^{z}C_{j}\frac{(n-n_{0})^{2}}{2},
 \label{KReq}
 \end{eqnarray}
 where ${\bf \sigma}\equiv (\sigma^{x},\sigma^{y},\sigma^{z})$ are the Pauli
 matrices, $J$ is the nearest-neighbor spin-spin interaction constant,
 $C_{j}$ and $n_{0}$ are the coefficient and minimum location
 of the parabolic kicking field, and $T_{0}$ is the kicking period.
 Below, we denote the $n=1$ ($n=N$) spin as the left (right) end of the chain.
 The constant field $B$ lifts the system degeneracy and the dynamics
 is restricted to a subspace with fixed total polarization $S_{z}$
 defined as
 \begin{eqnarray}
S_{z}\equiv \sum_{n=1}^{N}\sigma_{n}^{z}.
\end{eqnarray}
 Throughout this work we consider only the subspace of
$S_{z}=1$.

 Let $|{\bf m}\rangle$ be one of the basis states, with the $m$th spin up
 and all other spins down.
 The propagator for the time period $[jT_{0}-0^{+},(j+1)T_{0}-0^{+}]$
 is $\hat{V}(2JT_{0})\hat{U}(C_{j})$. Here $\hat{U}(C_{j})$
 represents the action due to the delta pulse, with
 \begin{eqnarray}
 \langle {\bf m}|\hat{U}(C_{j})|{\bf n}\rangle =\exp[-i(C_{j}/2)(n-n_{0})^{2}]\delta_{mn}.
\end{eqnarray} The
 term $\hat{V}$ stems
 from the evolution inherent in the  Heisenberg interaction.   An important recent
 study \cite{Bose06} has shown that,
 in the $N\rightarrow +\infty $ limit (and apart
 from some irrelevant phase)
 \begin{eqnarray}
 \langle {\bf m}|\hat{V}(2JT_{0})|{\bf n}\rangle \approx i^{(m-n)} J_{(m-n)}(2
 JT_{0}),
 \end{eqnarray}
 where $J_{(m-n)}$ is an ordinary Bessel function.
 The analytical behavior of $\hat{U}(C_{j})$ and $\hat{V}(2JT_{0})$
 is therefore completely in parallel with
 that associated with the propagator of the $\delta$-kicked rotor (DKR) (the
 best known model in quantum chaos \cite{casatibook}) with Hamiltonian
 \begin{eqnarray}
 H_{DKR}=(\hat{P}-P_{0})^{2}/2-K\cos(\theta)\sum_{j}\delta(t-j).
 \end{eqnarray}
 Indeed, in the representation of
 the basis states $|m\rangle\equiv \cos(m\theta)/\sqrt{\pi}$ and
 for an effective Planck  constant $\hbar$,
 the DKR propagator takes the familiar form $\hat{v}(k)\hat{u}(\hbar)$, with
 \begin{eqnarray}
\langle m|\hat{u}(\hbar)|n\rangle=\exp[-i(\hbar/2) (n-\tilde{n}_{0})^{2}]\delta_{mn},
\end{eqnarray}
 and  
 \begin{eqnarray}
 \langle m|\hat{v}(k)|n\rangle \approx i^{(m-n)} J_{(m-n)}(k)
 \end{eqnarray}
with ($k=K/\hbar$).
 Comparing these two systems, it is clear that  upon the mapping
 \begin{eqnarray}
|{\bf m}\rangle &\leftrightarrow & |m\rangle,  \\
2JT_{0}&\leftrightarrow & k,  \\
 n_{0}&\leftrightarrow & P_{0}/\hbar,\\
 C_{j}&\leftrightarrow& \hbar,
 \end{eqnarray}
 the many-body spin chain dynamics
 is
 mapped to that of DKR \cite{Bose06,boness,prosen98},  {\it i.e.},
 the motion of a spin wavepacket along the spin chain is mapped to 
DKR quantum diffusion dynamics in its $m$-space.
 Hence we can, whenever possible, shed light on
 the former
 by considering aspects of the latter, {\it e.g.},
 quantum resonance, Kolmogorov-Arnold-Moser (KAM) curves in phase space, etc.
More significantly for this work, as shown below, it allows us to use
tools from the control of quantum DKR dynamics
\cite{gongpre03,gongpre04,gongrev}
to manipulate states in
the spin chain. Further, this mapping between spin-chain and DKR allows
us to go beyond the parameter regime confined by the true DKR
(discussed below).

\section{Generation of Spin Wavepackets}
In the context of the wavepacket approach to quantum state transfer, we
now consider the first issue on spin wavepacket generation \cite{note1}.
Given a
small number of basis states that could be used for encoding the state of
a qubit, what initial
superposition states should be exploited to induce the creation of quantum
wavepackets with slow dispersion? Here this interesting question is
considered in the absence of an external field, where the system propagator is
given by $\hat{V}(2JT_{0})$.  Remarkably, the associated DKR
analogy now becomes a case of quantum resonance with $\hbar=4\pi$, with 
a propagator analogously given by
\begin{eqnarray}
\hat{v}(k=2JT_{0}) =\exp[i
(2JT_{0})\cos(\theta)].
\end{eqnarray}
Using this connection, the issue becomes
to find initial superposition states within a given small subspace, 
such that
the evolving  quantum state remains well
localized.  At first glance this ``localization" requirement
seems too demanding because
the main feature of quantum resonance dynamics is 
ballistic diffusion
in the DKR $m$-space. However, as we have discovered,
this can still be obtained by maximizing
a diffusion rate of DKR.
Qualitatively speaking,
for superposition states maximizing a quadratic diffusion rate
introduced below,
 the ensuing dynamics
 will push outwards as much as possible the excitation profile
 in the $m$-space,
 thus generating two well-separated wavepackets with almost zero amplitude
 in between.

 \begin{figure}
 \begin{center}
 \epsfig{file=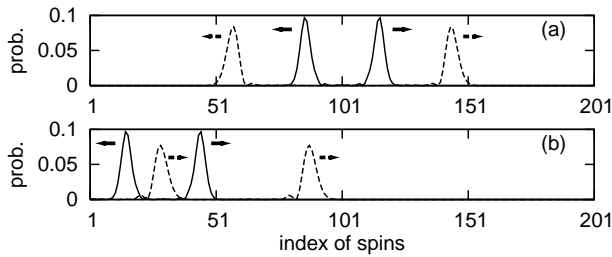,width=8.6cm}
 \end{center}
 \caption{Emergence of a wavepacket pair in a Heisenberg chain of $201$
 spins, shown with the projection probability of the many-body wavefunction onto
 basis states $|{\bf m}\rangle$.
 The initial condition is a superposition state $|\psi_{m_{0}}\rangle$
given by Eq.  (15), for $m_{0}=101$ in (a) and $m_{0}=30$ in (b).
Th system wavefunction then evolves, with its shape
given by the solid lines at time $t_{1}$ with
 $2J t_{1}=15$, and by the dashed lines 
 after an additional period $t_{2}$, with $2J t_{2}=30$ in (a)
 and $2Jt_{2}=45$ in (b).
 The arrows show the travel direction of the wavepackets.}
 \end{figure}

 Quantitatively, let us first define the diffusion rate operator as
 \begin{eqnarray}
 \hat{D}=\lim_{t \rightarrow +\infty}\frac{\hat{E}(t)-\hat{E}(0)}{t^2},
 \end{eqnarray}
 where $\hat{E}(t)$ is
 the energy operator for the free rotor in the Heisenberg representation.
 For the quantum resonance case considered here  one obtains
 \begin{eqnarray}
 \hat{D}= A\sin^{2}(\theta),
 \end{eqnarray}
 where
 $A$ is  a constant.  Note that $\hat{D}$ only couples states $|m\rangle$ of the same parity.
 Consider now a sample case where
 a superposition state
 \begin{eqnarray}
 |\psi_{m_{0}}\rangle=\sum_{n=-2}^{n=2}\beta_{2n}|m_{0}+2n\rangle
 \end{eqnarray}
 is
 exploited
 to encode a qubit state, {\it i.e.},
 only five basis
 states are used here.
 The state with  the largest diffusion rate, denoted $D$,
 is simply given by  the eigenfunction of
 $\hat{D}$ in the subspace of
 $|m_{0}+2n\rangle$ ($-2 \leq n \leq 2$)
 with the largest eigenvalue.
 In particular, if these basis states do not involve state
 $|0\rangle$, then the maximized $D$ is attained if
 $\beta_{0}= 0.577 $, $\beta_{-2}=\beta_{+2}=-0.5 $,
 $\beta_{-4}=\beta_{+4}=0.289 $.

 The significance of such an initial superposition state
 with maximized $D$ is demonstrated in
 Fig. 1(a),  with
 $m_{0}=101$ for a $201$-spin chain.
 In particular,
 a well-separated wavepacket pair is seen to quickly emerge,
  and its dispersion after its emergence
  is impressively slow in the absence of any external static fields.
  Note that, unlike the accelerator mode approach
    proposed in Ref. \cite{Bose06},
  the wavepacket pair is created by the system dynamics itself. Note also
  that 
  the excitation amplitudes
  between the two wavepackets are surprisingly  small.
  Because the total polarization here is fixed, 
  a well-separated wavepacket pair
  directly
  indicate quantum entanglement between well-separated parts of the spin chain,
  and their travel in
  opposite directions
  distributes information or entanglement to both ends of the spin chain.
  Further results (not shown) indicate that
  in the case of
  minimized $D$ or an arbitrarily chosen initial state,
  the system dynamics generically  creates a quickly delocalizing
  state (note also that even in chaotic cases
  different initial superposition states may also lead
  to dramatic differences in the ensuing  quantum diffusion dynamics 
  \cite{gongprl}).
  These further demonstrate
  the important role of
  an initial superposition state with a maximized diffusion rate in encoding
  the quantum information. Certainly, if more basis states are
  allowed in encoding the quantum information, then wavepacket pairs
  with even slower dispersion can be created with the same approach.

  It is also desirable to be able to create a well-separated wavepacket
  pair that transfers information to
  a common end of the spin chain.
  For example, if two wavepackets with identical shape 
  can be created, then one of them may be analogous
  to a ``backup" copy as the other
  is being transferred and received first. Note that this
  possibility is not in violation of the quantum no-cloning theorem,
  because here the two wavepackets do not independently describe
  the quantum state of the involved spins. Rather, the two wavepackets
  describe the entanglement between two particular
  sections of the spin chain.

  The creation of such a wavepacket pair is achieved here
  by going beyond the kicked rotor perspective and exploiting
  the boundary effect
  associated with the spin chain. That is, we
  apply the above scenario, but with
  the initial encoding state $|\psi_{m_{0}}\rangle$ located at
  $m_{0}<N/2$, and with the requirement that no
  information receiver presents at the left ($n=1$) end.
  To be more specific, consider  a sample result
  shown in Fig. 1(b), with $m_{0}=30$.
  The wavepacket creation dynamics in the early
  stage is seen to be analogous to the case of Fig. 1(a).  Sometime later,
  the generated wavepacket moving to the left hits the boundary and gets
  reflected. As demonstrated in Fig. 1(b),
  this then creates a pair of wavepackets where both of the members of the 
pair are moving to
  the right, with their shape indistinguishable from one another,
  with almost zero excitation in-between,
  and a peak-to-peak distance of   $2(m_{0}-1)$ spins.

\section{Stopping and relaunching Wavepacket Propagation}
In this section, taking the wavepackets obtained in the previous section
as examples, we shall consider an independent issue, namely,
  stopping and
  relaunching the quantum state transfer along a spin chain.
  A parabolic magnetic field
  [see Eq. (\ref{KReq})] is proposed as the control field with
  a simple global feature, and we aim to achieve
  our control objective with an always-on Heisenberg interaction.

  Thanks to the DKR analogy discussed above,  the problem is now converted
  into the question of how to
  stop and successfully restart the transport process in the DKR $m$-space.
We do so below by exploiting control features of the DKR system.
Two features are relevant, one classical and one quantal.

  The classical  basis of our control scenario arises by
  exploiting the phase space KAM curves of the underlying classical dynamics.
  That is, if the kicking magnetic field is sufficiently
frequent that the chaoticity parameter $|2JT_{0}C_{j}|$ in the classical
DKR is sufficiently small, the underlying
  classical dynamics will be mainly integrable and the associated KAM curves
  will present strong barriers to
  the quantum transport in the $m$-space.
  Because KAM curves will be almost everywhere,
  these classical structures can effectively stop the travel of arbitrary
  and unknown quantum wavepackets.

  Figure 2 displays the fate of the moving wavepacket pair shown in Fig. 1(a) (solid lines)
  after a kicking parabolic field is introduced.
As is clearly seen in Fig. 2(a) and Fig. 2(b),
  the transfer of the wavepacket pair to both ends of the spin chain is stopped.
  This dynamical effect can also be understood as a type
  of quantum Zeno effect achieved by frequently applying 
(but far from infinitely fast) external pulses to a system.

  However, 
  although the wavepacket pair in Figs 2(a) and 2(b)
has stopped moving, its internal structure is
  seen to be changing in a subtle manner. This indicates that
  evolution of the quantum phases
  characterizing the stopped wavepackets
  is still not frozen.  This fact turns out to be disastrous when the kicking
  field is turned off in order to relaunch the state transfer.
  For example, Fig. 2(c) displays the wavefunction after
  the kicking field has been off
  for a period of $t_{2}$ with $2Jt_{2}=30$: the background fluctuation
  is greatly increased, and the main peaks of the wavepacket pair
  do not move further.

  \begin{figure}
  \begin{center}
  \epsfig{file=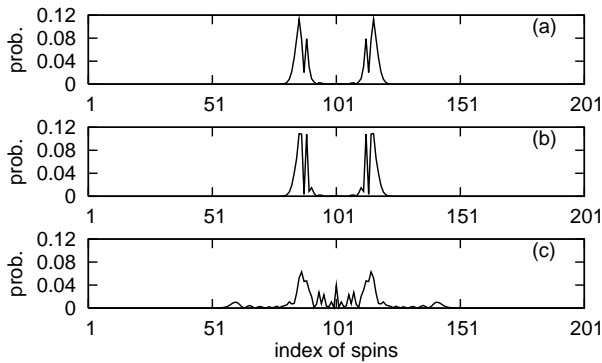,width=8.6cm}
  \end{center}
  \caption{Stopping of quantum state transfer
  in a spin chain by a parabolic kicking field,
  with $C_{j}=0.5$.
  The initial state, the meaning of the wavefunction profile,
  and the spin chain used for calculations are
  the same as in Fig. 1(a). $2JT_{0}=0.25$, and the first
  $\delta$-kick comes at time $t_{1}$ with $2Jt_{1}=15$.
  (a) and (b) are the results after 100 and 200 kicks.
  The kicking field is then turned off and
  (c) displays the state after an additional period $t_{2}$
  with $2Jt_{2}=30$.}
  \end{figure}

Hence, using KAM curves alone, which is  a purely classical
  control mechanism,
  does not offer a satisfactory means of stopping the wavepacket.
  To improve the control one must compensate
  for the quantum phases that are accumulated during the stopping process.
  This phase accumulation is due to the spin-spin
  interaction as well as the kicking field.

  Consider then an important observation made in our previous work on
  the quantum control of
  DKR dynamics \cite{gongpre03,gongrev}, {\it i.e.}, 
  \begin{eqnarray}
  \langle m| \hat{v}(k) |n\rangle 
& = & \frac{1}{\pi}\int^{2\pi}_{0}\cos(m\theta)
  \exp[ik\cos(\theta)] \cos(n\theta)d\theta \nonumber \\
  & = & \langle m|(-1)^{m} \hat{v}(-k) (-1)^{n} |n\rangle \nonumber \\
  & = & \langle m|\hat{u}(2\pi) \hat{v}(-k) \hat{u}(2\pi) |n\rangle.
  \label{vuv}
  \end{eqnarray}
  The first line in Eq. (\ref{vuv}) holds by definition, the second line
  becomes obvious if the integration variable $\theta$ is changed
    to $\theta+\pi$, and the last line is obtained by use of the definition of
        $\hat{u}(\hbar)$.  Equation (\ref{vuv}) hence proves
\begin{eqnarray}
 \hat{v}(k)=\hat{u}(2\pi) \hat{v}(-k) \hat{u}(2\pi).
\end{eqnarray}
  Returning to a finite spin chain system, this result indicates that
  \begin{eqnarray}
  \hat{V}(2JT_{0})\approx \hat{U}(2\pi) \hat{V}(-2JT_{0})
  \hat{U}(2\pi).
  \end{eqnarray}
  That is, the sign of the intrinsic
  interaction constant $J$ can be effectively reversed if
  we apply two parabolic $\delta$-kicks of  particular strength.
  As such, it becomes possible to compensate for
  the quantum phase evolution inherent in the spin chain.  As to
  the quantum phases induced by the kicking field, they can also be
  compensated for
  by considering kicking fields with the sign of $C_{j}$ reversed.

  \begin{table}
  \caption{The $j$-dependence of $C_{j}$ [see Eq. (1)] in an explicitly designed pulse
  sequence for the stopping of arbitrary wavepackets for a period of
  $2M$ kicks. $C$ is a constant discussed in the text. Note that some
  system parameters used here
  are beyond what is allowed in a true kicked rotor system.}
  \begin{center}
  \begin{tabular}{|c|c|c|c|c|c|}
  \hline
  $\ j \ $ &  $1$ & $(1, M]\ $ &\  $M+1$\ &\ $(M+1,2M]\ $ &\ $2M+1$\   \\
  \hline
  \ $C_{j}\ $ & $C/2+2\pi $ &  $C$ &  $2\pi$ & $-C$ & $-C/2$ \\  \hline
  \end{tabular}
  \end{center}
  \end{table}

  Given these considerations we present in Table I
  an explicitly designed special pulse
  sequence that can relaunch stopped wavepackets with perfection.
  For this special pulse sequence,
  the KAM curves associated with small $|2JT_{0}C|$
  still play a key role because they
  directly prevent the state transfer, in the same manner as demonstrated
  in Fig. 2.
  What is remarkable now is the total time evolution operator
  associated with the entire stopping process.
  In terms of the DKR analogy, this operator can be written as (after some
  manipulation)
  \begin{eqnarray}
  && \left[\hat{u}(-C/2)\hat{v}(k)\hat{u}(-C/2)\right]^{M}
  \left[\hat{u}(C'/2)\hat{v}(k)\hat{u}(C'/2)\right]^{M}
   \nonumber \\
   & =& \left[\hat{u}(-C/2)\hat{v}(k)\hat{u}(-C/2)\right]^{M}
   \left[\hat{u}(C/2)\hat{v}(-k)\hat{u}(C/2)\right]^{M} \nonumber \\
    & = & \left[\hat{u}(-C/2)\hat{v}(k)\hat{u}(-C/2)\right]^{M'}
     \left[\hat{u}(C/2)\hat{v}(-k)\hat{u}(C/2)\right]^{M'} \nonumber \\
      &=& \cdots = 1,
       \label{phaseeq}
       \end{eqnarray}
       where $C'=C+4\pi$, $M'=M-1$. In obtaining
       Eq. (\ref{phaseeq}) we have used
       \begin{eqnarray}
       \hat{u}(C)=\hat{u}(C/2+2\pi)\hat{u}(C/2+2\pi)
       \end{eqnarray}
       and Eq. (\ref{vuv}). Equation (\ref{phaseeq}) proves that
       at the end of the stopping time
       all properties characteristic of an {\it unknown} quantum wavepacket 
       can be
       exactly restored. 
       This exact rephasing indicates that
       the dynamical evolution 
       associated with the second $M/2$ kicks, in addition to 
       offering
       a dynamical barrier to stop the quantum transport,
       precisely
       reverses the evolution associated with the first $M/2$ kicks.
       As such, the stopping is entirely {\it nondestructive}, as long
       as the system is not subject
       to noise effects during the stopping process.
       Evidently then,
       wavepacket-assisted information transfer can be perfectly relaunched
       as the kicking field is turned off.
       This theoretical result applies exactly to
       an infinitely long spin chain. But fortunately,
       as also demonstrated below, it
       applies extremely well to a finite-length chain.
       Note also
       that the designed pulse sequence in Table I
       is a significant extension beyond a true DKR system because
       both positive and negative
        ``$\hbar$" ($\hbar \leftrightarrow C_{j}$) are exploited here.

       \begin{figure}
       \begin{center}
       \epsfig{file=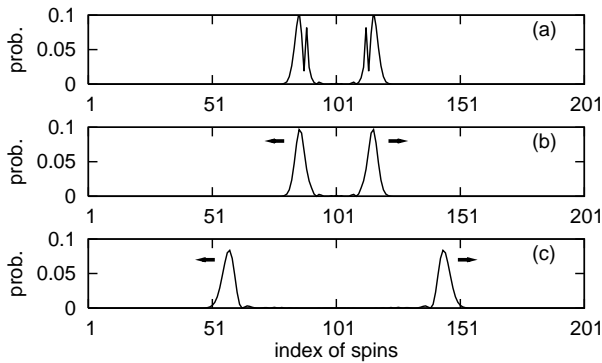,width=8.6cm}
       \end{center}
       \caption{
       Same as in Fig. 2, but for an explicitly designed
           pulse sequence given in Table I, with $C=0.5$, $M=100$.
	    The transfer of
         the wavepacket pair along the spin chain is stopped
          for $2M$ kicks.
      Panels (a) and (b) are the results after 100 and 200 kicks.
       The result in (c) shows that the state transfer is
       indeed  relaunched with perfection.}
       \end{figure}

       In parallel with Fig. 2, Fig. 3 displays a computational example using
       the pulse sequence given in Table I.
       As in Fig. 2, right before the first kick the quantum state
       is the wavepacket pair described by the solid lines in Fig. 1(a).
       Figure 3(a) confirms that the
       wavepacket pair
       is not moving as the special pulse sequence is on.
       Figure 3(b) shows that at the end of the stopping period,
       we have restored the initial condition [compare Fig. 3(b) with
       the solid lines in Fig. 1(a)].
       The restoration fidelity in the numerical 
       calculations for a $201$-spin chain
       is found to be higher than
       $[1- 10^{-13}]$. 
       The kicking field is then turned off.
       As expected, quantum state transfer is relaunched and the wavepackets
       continue their journey, with slow dispersion,
       towards
       both ends of the spin chain [Fig. 3(c)].
       Indeed, results in Fig. 3(c) are indistinguishable from
       the dashed lines in Fig. 1(a).
       
 The control scenario proposed in this work can also lead to other very interesting approaches to the manipulation
 of quantum entanglement dynamics of a spin chain.
 Here we briefly discuss three possibilities.
 First, by modifying
       the kicking field
       profile we can choose to stop only one component
       of a  wavepacket pair, e.g., of the pair shown
       in Fig. 1(b) with dashed lines, thereby
       offering an interesting  method of
       tuning the time delay between two
       wavepackets moving in the same direction.  This then offers a means of controlling the distance 
       between two entangled parts of the spin chain.
       Second, because
        the sign of the intrinsic spin-spin
	 interaction constant $J$ can be effectively reversed if
   we apply $\delta$-kicks of  particular strength,
       it can be easily shown that one can
       bounce back an arbitrary and unknown moving wavepacket to
       the sender at a time of our choosing.
       Third, by controlling the time delay between the two wavepackets and/or
       taking advantage of the feasibility of time-reversal, 
       we may also recombine two localized wavepackets at a location
       different from that of the initial state. This recombination dynamics
       resembles that of
       a double-slit experiment, thereby generating interesting interference patterns along
       the spin chain.  Such kind of interference patterns of
       spin excitations, and their fate under a variety of circumstances,
      may work as a novel interferometer for
      fundamental studies in quantum physics.

\section{conclusion}
       To conclude, based on a mapping between a kicked spin
       chain and the delta-kicked rotor system \cite{Bose06}, 
       we have shown that
       previous quantum control results in
       the delta-kicked rotor system \cite{gongpre03,gongpre04,gongrev,gongprl}
       can be applied to the control of
       spin wavepacket propagation and hence the control of propagating
       quantum information
       encoded in wavepackets.
       Specifically, we have proposed a simple approach to wavepacket creation
       in a Heisenberg spin chain and demonstrated the
       possibility of stopping and relaunching
       information transfer without individually 
       addressing spins or turning off
       spin-spin interactions.
        Several interesting
       applications of this work in manipulating the dynamics of a spin chain
       are also discussed.
       The results indicate that many insights from
       the quantum chaos research can be very useful for
       quantum information science. 
       This work also
       adds more support to 
       the use of spin chains as quantum wires, and might be
       useful in designing
       new quantum computation algorithms with an always-on qubit-qubit
       interaction \cite{Bose03b,Twamley06}.
       Extensions to other types of spin chains are under consideration.

{\bf Acknowledgments}:
The authors thank one of the referees for thought-provoking comments.
J.G. is supported by the start-up funding
(WBS No. R-144-050-193-101 and No. R-144-050-193-133),
National University of Singapore,
and the ``YIA" funding (WBS No. R-144-000-195-123) from the office of 
Deputy President (Research \& Technology),
National
University of Singapore. P.B. was
supported by the Natural Sciences
       and Engineering Research Council of Canada.
       J.G. thanks Prof. Berthold-Georg  Englert for an interesting 
       comment on this work,
       and Dario Poletti and Wenge Wang for
       critical reading of this manuscript.


\begin{thebibliography}{100}
\bibitem{note1}Within the framework of the wavepacket approach to
	quantum state transfer, the decoding of the transported wavepackets 
	is assumed to be doable and is not discussed in this work.
\end{thebibliography}
        \end{document}